**Excitation of coupled phononic frequency combs via two-mode parametric three-wave mixing**

Authors: Adarsh Ganesan[1], Cuong Do[1], Ashwin Seshia[1]

[1.] Nanoscience Centre, University of Cambridge, Cambridge, UK

**This paper builds on the recent demonstration of three-wave mixing based phononic frequency comb. Here, in this process, an intrinsic coupling between the drive and resonant frequency leads to a frequency comb of spacing corresponding to the separation between drive and resonant frequency. Now, in this paper, we experimentally demonstrate the possibility to further excite multiple frequency combs with the same external drive through its coupling with other identical devices. In addition, we also experimentally identify interesting features associated with such a frequency comb generation process.**

This paper builds upon the recent experimental demonstration of the phononic counterpart of optical frequency combs in a micromechanical vibratory device [1]. The generation of these phononic frequency combs is described by a nonlinear three-wave mixing process as theoretically outlined in [2]. Through this pathway, a single drive tone results in the equidistant and phase-coherent spectral lines corresponding to the frequency comb and the spacing of such a frequency comb corresponds to the separation between the drive and resonant frequency. In this paper, we experimentally demonstrate the excitation of frequency combs of two different characteristic spacing utilizing a single drive tone.

For this demonstration, we consider an experimental system of two-coupled Si-based micromechanical free-free beam structures of dimensions $1100\ \mu m$ X $350\ \mu m$ X $11\ \mu m$. The mechanical coupler has dimensions 20 $\mu m$ X 2 $\mu m$ X 11 $\mu m$ beam (Figure 1A). The device also consists of the $0.5\ \mu m$ thick AlN and $1\ \mu m$ thick Al layers above the Si surface for its piezoelectric actuation. To preserve the uniformity, the AlN and Al patterns of the two free-free beam microstructures are identical. The signal derived from Agilent 335ARB1U is applied on the device through the Al electrodes and the resulting electrical response corresponding to the mechanical motion of the device is probed using Agilent infiniium 54830B DSO. For the independent validation, the resonant motion of the device is also monitored using Laser Doppler Vibrometer (LDV).

Figure 1B shows the output spectrum when an electrical signal of $S_{in}(f_d = 3.86\ MHz) = 23\ dBm$ is applied. In this figure, there exists two thick spectral features closer to $f_d$ and $\frac{f_d}{2}$. The zoomed-in views indicate the correspondence of these features to the frequency combs. To explore these frequency combs further, the experiments were carried out at wide-ranging drive power levels for the same drive frequency $f_d = 3.86\ MHz$. The electrical spectra $S_{out}$ for $S_{in}(f_d = 3.86\ MHz)$ ranging from $-10\ dBm$ to $23.5\ dBm$ are presented in the condensed figure 1C. For the drive levels below $10\ dBm$, only the drive tone is observed. However, the further increase in $S_{in}$ leads to a frequency comb. It is now surprising to note the drastic shift in the spacing of the frequency comb above the critical $S_{in}^* = 16.5\ dBm$. In order to conclude whether this observed shift corresponds to an auxiliary nonlinear pathway associated with the generation of the same frequency comb 1 or results from the transition to a different frequency comb altogether, the nature of the frequency combs around the parametrically excited sub-harmonic tone is investigated.

At a low drive level of $-5\ dBm$, the parametric resonance is absent. Hence, only the drive tone is observed (Figure 2A). However, when the drive level is further increased to $5\ dBm$, in addition to the drive tone, there is also an excitation of sub-harmonic tone corresponding to the parametric resonance (Figure 2B). Despite this, the frequency comb is not formed at this drive condition. However, for the drive levels $15\ dBm$ and $23\ dBm$, both the frequency combs and parametric resonance are observed (Figures 2C and 2D). The different drive power level thresholds associated with the parametric resonance and frequency combs indicate that the parametric resonance is only a necessary but not sufficient condition for the frequency comb formation. The figures 2C and 2D also present interesting qualitative differences in the frequency combs formed at the drive levels $15\ dBm$ and $23\ dBm$. While the tone $\frac{f_d}{2}$ is observed in the frequency combs at $15\ dBm$, such a tone is absent at $23\ dBm$. Hence, in addition to the increased frequency spacing of combs at $23\ dBm$ as compared to those at $15\ dBm$, there is also a fundamental difference in the nature of these combs. So, the observed drastic frequency shift above the critical $S_{in}^* = 16.5$ in the figure 1C cannot simply be due to an additional nonlinear process associated with the generation of the same frequency comb. However, it has to be explained by the transition from one class of frequency combs to another. Such classes of frequency combs can be qualitatively described by $f_d \pm n(f_d - \tilde{f}_1); \frac{f_d}{2} \pm n(f_d - \tilde{f}_1)$ and $f_d \pm n(f_d - \tilde{f}_2); \frac{\tilde{f}_2}{2} \pm n(f_d - \tilde{f}_2)$ respectively. Here, $\tilde{f}_1$ and $\tilde{f}_2$ correspond to the two different re-normalized resonant frequencies.

To understand these experimental observations in the context of previous demonstration of frequency comb in [1], the following dynamics is considered.

$$\ddot{Q}_i = -\omega_i^2 Q_i - 2\zeta_i \omega_i \dot{Q}_i + \sum_{\tau_1=1}^{2}\sum_{\tau_2=1}^{2} \alpha_{\tau_1\tau_2}\, Q_{\tau_1}Q_{\tau_2} + \sum_{\tau_1=1}^{2}\sum_{\tau_2=1}^{2}\sum_{\tau_3=1}^{2} \beta_{\tau_1\tau_2\tau_3}\, Q_{\tau_1}Q_{\tau_2}Q_{\tau_3} \qquad (1)$$
$$+ P\cos(\omega_d t)$$

where $P$ is the drive level, $\alpha$ and $\beta$ are quadratic coupling coefficients and $\omega_{i=1,2}$ and $\zeta_{i=1,2}$ are natural frequencies and damping coefficients of modes $i = 1,2$ respectively.

Here, when this coupled system of 2 modes is driven closer to $\omega_1$, the mode 1 gets directly excited. With an increased power level $P$, the modal displacement $Q_1$ may also get high enough to trigger parametric excitation of mode 2 through $Q_1Q_2$ nonlinearity. Such parametrically excited tone is expected to have a frequency $\frac{\omega_d}{2}$ and is closer the resonant frequency of mode 2: $\omega_2$. However, the recent experiments on the two-mode three wave mixing based frequency comb [1] have indicated the possibility for $\frac{\omega_1}{2}$ excitation instead of $\frac{\omega_d}{2}$. This particularly occurred when $\omega_d$ is set outside the dispersion band i.e. $|\omega_d - \omega_1| > \delta$. Once the parametric excitation of $\frac{\omega_1}{2}$ is introduced, the frequency combs $\omega_d \pm n(\omega_d - \omega_1); \frac{\omega_1}{2} \pm n(\omega_d - \omega_1)$ are generated through the high-order nonlinear mixing processes.

$$\ddot{Q}_i = -\omega_i^2 Q_i - 2\zeta_i\omega_i \dot{Q}_i + \sum_{\tau_1=1}^{3}\sum_{\tau_2=1}^{3} \alpha_{\tau_1\tau_2} Q_{\tau_1}Q_{\tau_2} + \sum_{\tau_1=1}^{3}\sum_{\tau_2=1}^{3}\sum_{\tau_3=1}^{3} \beta_{\tau_1\tau_2\tau_3} Q_{\tau_1}Q_{\tau_2}Q_{\tau_3}$$
$$+ P\cos(\omega_d t); i = 1,2,3 \qquad (2)$$

Now, we turn to a system of three coupled modes and the modified dynamics is presented in equation 2. Let us consider the case where the drive frequency $\omega_d$ is simultaneously closer to the resonant frequencies of two modes: $\omega_1$ and $\omega_2$ (Supplementary Figure S1). Further, $\omega_d$ is also closer to twice the resonant frequency of a third mode $2\omega_3$. In this case, the modes 1 and 2 are directly excited at first. For high drive power levels of $P$, the modal displacements $Q_1$ and $Q_2$ may also get high enough to also independently parametrically excite mode 3 through $Q_1Q_3$ and $Q_2Q_3$ nonlinearities. Based on the current understanding, the frequency of this tone is expected to be $\frac{\omega_d}{2}$. However, due to the nonlinear three-wave mixing processes, deviations from this frequency may also be encountered similar to the case presented before. While such deviations are possible through the dynamical equation (2) as demonstrated numerically in [2], the analytics describing them as a function of multiple system parameters are not known yet. Qualitatively, since $\omega_d$ is closer to both $\omega_1$ and $\omega_2$, the parametric excitation of tones at either $\frac{\omega_1}{2}$ or $\frac{\omega_2}{2}$ or even $\frac{\omega_d}{2}$ may be possible. Following the excitation of $\frac{\omega_1}{2}$ or $\frac{\omega_2}{2}$, the high-order mixing may result in the frequency combs $\omega_d \pm n(\omega_d - \omega_1); \frac{\omega_1}{2} \pm n(\omega_d - \omega_1)$ or $\omega_d \pm n(\omega_d - \omega_2); \frac{\omega_2}{2} \pm n(\omega_d - \omega_2)$ respectively. In contrast, when $\frac{\omega_d}{2}$ is excited, we may not expect a frequency comb based on our current understanding of frequency comb. However, our experiments show otherwise. That is, the frequency combs $\omega_d \pm n(\omega_d - \omega_2); \frac{\omega_d}{2} \pm n(\omega_d - \omega_2)$ are observed at certain drive conditions (Figure 2B). To explain the emergence of such frequency combs, we sketch an underpinning pathway as follows. The parametrically excite sub-harmonic tone $\frac{\omega_d}{2}$ may excite $\omega_2$ through the parametric back-action. This may in turn lead to the observed nature of frequency combs $\omega_d \pm n(\omega_d - \omega_2); \frac{\omega_d}{2} \pm n(\omega_d - \omega_2)$. Also, the frequencies $\omega_1$ or $\omega_2$ may also be drive power level dependent. Hence, we replace $\omega_1$ and $\omega_2$ by the re-normalized frequencies $\widetilde{\omega}_1$ and $\widetilde{\omega}_2$ respectively. We have thus qualitatively argued the possibility for the multiple classes of frequency combs in the context of the equation 2. Now, we experimentally map the regimes specific to these classes of frequency combs at different drive conditions.

For a drive frequency $f_d = \frac{\omega_d}{2\pi} = 3.86\ MHz$, we had already mapped the nature of resonances for a range of power levels $-10 - 23.5\ dBm$ (Figure 1B). This revealed the transition from $\omega_d$ (No parametric resonance; No frequency comb) to $\omega_d; \frac{\omega_d}{2}$ (No frequency comb) to $\omega_d \pm n(\omega_d - \widetilde{\omega}_1); \frac{\omega_d}{2} \pm n(\omega_d - \widetilde{\omega}_1)$ (Frequency comb 1) to $\omega_d \pm n(\omega_d - \widetilde{\omega}_2); \frac{\widetilde{\omega}_2}{2} \pm n(\omega_d - \widetilde{\omega}_2)$ (Frequency comb 2). Now, we chart out such regimes for different drive frequencies and power levels. As shown in the figure 3, the regimes specific to the excitation of frequency comb 1 and 2 are non-standard and are not merely characterised by mere parametric resonance threshold. For instance, for the drive frequencies $3.86\ MHz < f_d < 3.863\ MHz$, the comb and parametric excitation thresholds are not the same (Figure 3). Hence, there exist regimes where the parametric excitation takes place without the frequency comb formation. Also, for $3.855\ MHz < f_d < 3.857\ MHz$, the frequency comb is formed only for a specific range of drive power levels (Figures 3b1 and 3b2). Interestingly, the regimes corresponding to the frequency combs 1 and 2 are well-bounded. Despite this, they do not completely relate to the previous understanding of frequency comb regimes. Previously, the frequency combs were only observed outside the dispersion band. In contrast, the current experiments show the generation of frequency combs even within the dispersion band.

As briefly mentioned before, the frequency comb spacing is drive level dependent and the figure 4 presents this dependence. While the spacing increases with the drive power level, it is nearly constant with the drive frequency at a specific drive power level (Figures 4A and 4C). This is similar to the case presented in the previous work (cf. Figure 3 in [1]). However, in contrast to the linear increase with the drive power level (cf. Figure 3 in [1]), the dependences in the current experiments are quadratic (Figure 4B) and cubic (Figure 4D).

In summary, the excitation of two coupled frequency combs using two-mode parametric three-wave mixing has been observed in a specific system of two mechanically coupled micromechanical resonators. The fundamental nature of the component frequency combs is different. While the tone at half the drive frequency $\frac{f_d}{2}$ is observed in one frequency comb, it is absent in the other. Also, the spacing and its nonlinear drive level dependence are different for such frequency combs. There also exist mutually exclusive well-bounded regimes for each component frequency combs and such regimes are not merely described by the parametric resonance threshold. In future, multiple experiments with this micromechanical resonator platform are warranted to attain new insights into the nonlinear physics and also to arrive at the rigorous analytical descriptions modelling the precise nature of frequency combs.

**Acknowledgements**

Funding from the Cambridge Trusts is gratefully acknowledged.

**References**

[1] A. Ganesan, C. Do, and A. Seshia, "Phononic Frequency Comb via Intrinsic Three-Wave Mixing," *Physical Review Letters,* vol. 118, p. 033903, 2017.

[2] L. S. Cao, D. X. Qi, R. W. Peng, M. Wang, and P. Schmelcher, "Phononic Frequency Combs through Nonlinear Resonances," *Physical Review Letters,* vol. 112, p. 075505, 2014.

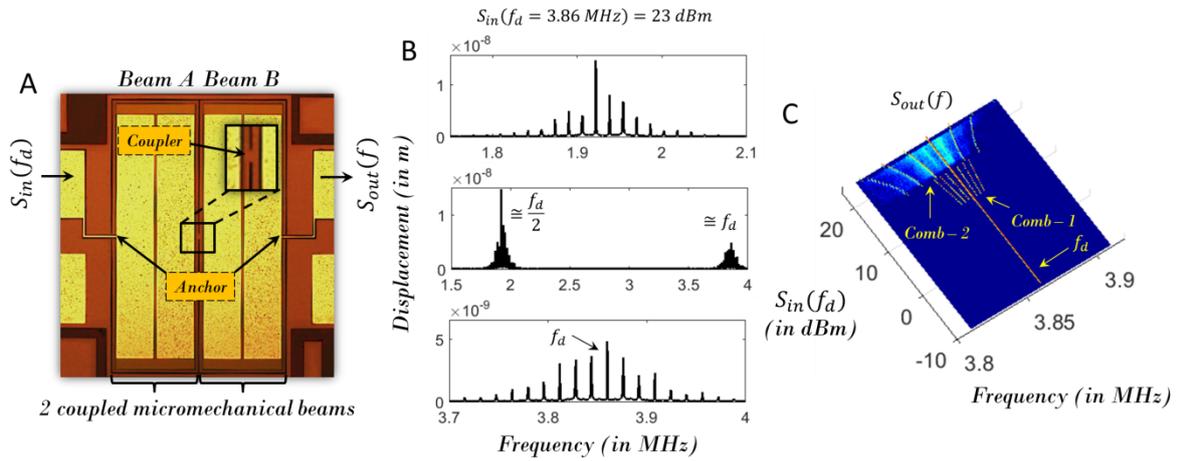

Figure 1: **Observation of coupled phononic frequency combs via two-mode three-wave mixing.** A: An electrical signal $S_{in}(f_d = 3.857\ MHz)$ is applied on a mechanically coupled free-free beam microstructure; B: The frequency spectrum of the output electrical signal $S_{out}$ for $S_{in}(f_d = 3.857\ MHz) = 23\ dBm$ (measured using Laser Doppler Vibrometry). Top: The zoomed-in view of the spectrum around $f_d/2$. Middle: The zoomed-out view. Bottom: The zoomed-in view of the spectrum around $f_d$; C: The spectral maps of the output electrical signal $S_{out}$ for different drive conditions $S_{in}(f_d = 3.86\ MHz) = -10 - 23.5\ dBm$ (measured using electrical spectrum analyser).

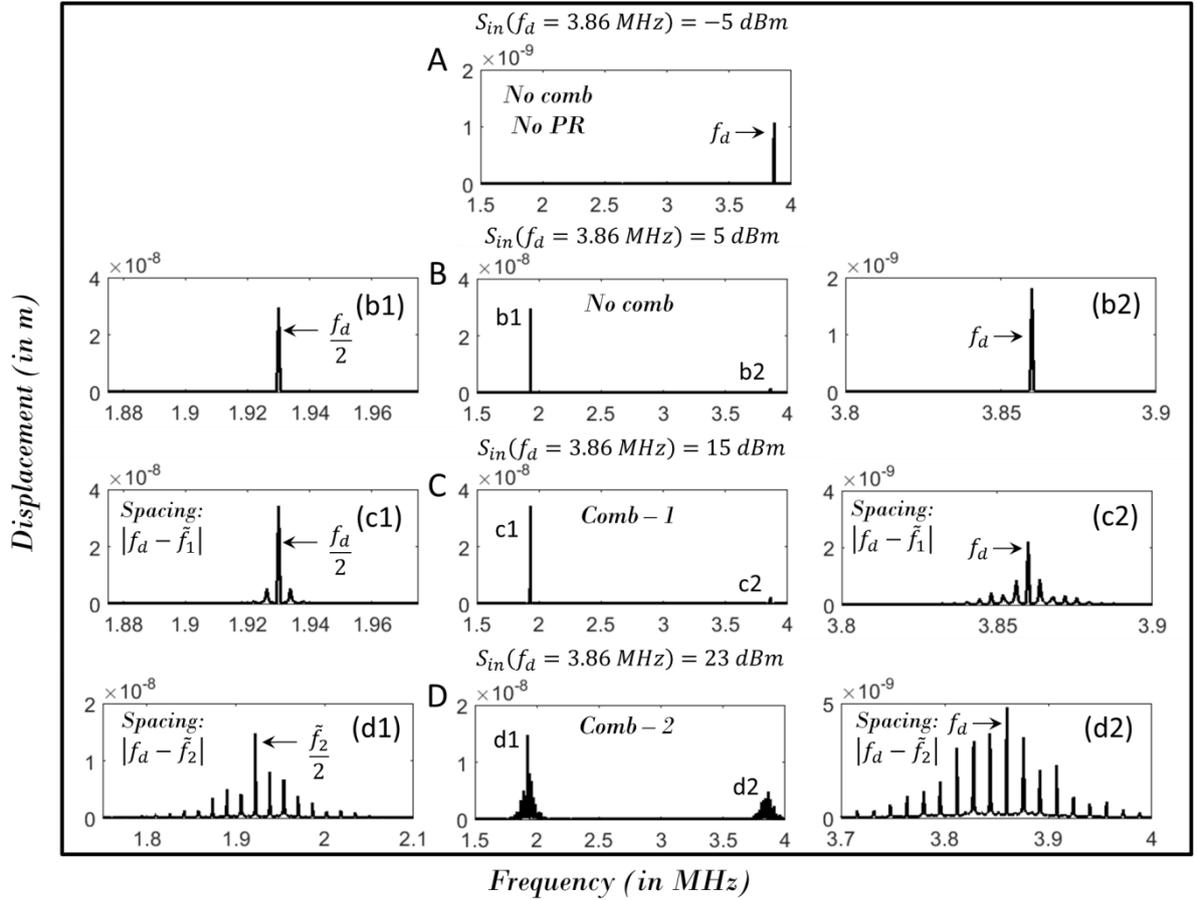

Figure 2: **Observation of coupled phononic frequency combs via two-mode three-wave mixing (Contd.).** A-D: The frequency spectra of the output electrical signal $S_{out}$ for $S_{in}(f_d = 3.857\ MHz) = -5, 5, 15\ \&\ 23\ dBm$ respectively (measured using Laser Doppler Vibrometry); b1, c1 & d1: The zoomed-in views of the spectra B, C and D around $f_d/2$ respectively; b2, c2 & d2: The zoomed-in views of the spectra B, C and D around $f_d$ respectively.

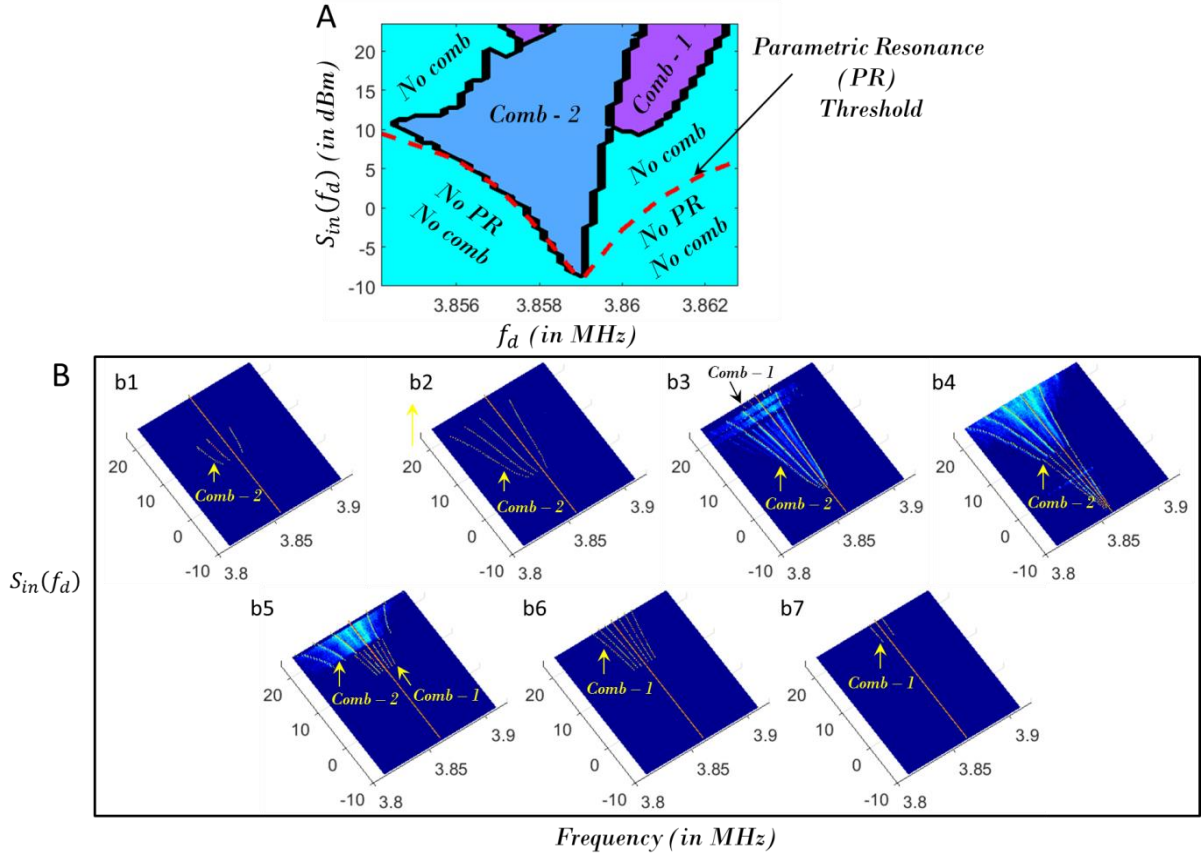

Figure 3: **Regimes of phononic frequency combs.** A: The regimes of phononic frequency combs for a range of drive frequencies $f_d = 3.8544 - 3.8626\ MHz$ and drive power levels $S_{in} = -10 - 23.5\ dBm$; B: b1-b7: The drive power level dependent output spectra $S_{out}$ for different drive frequencies $f_d = 3.856, 3.857, 3.858, 3.859, 3.86, 3.861\ \&\ 3.862\ MHz$.

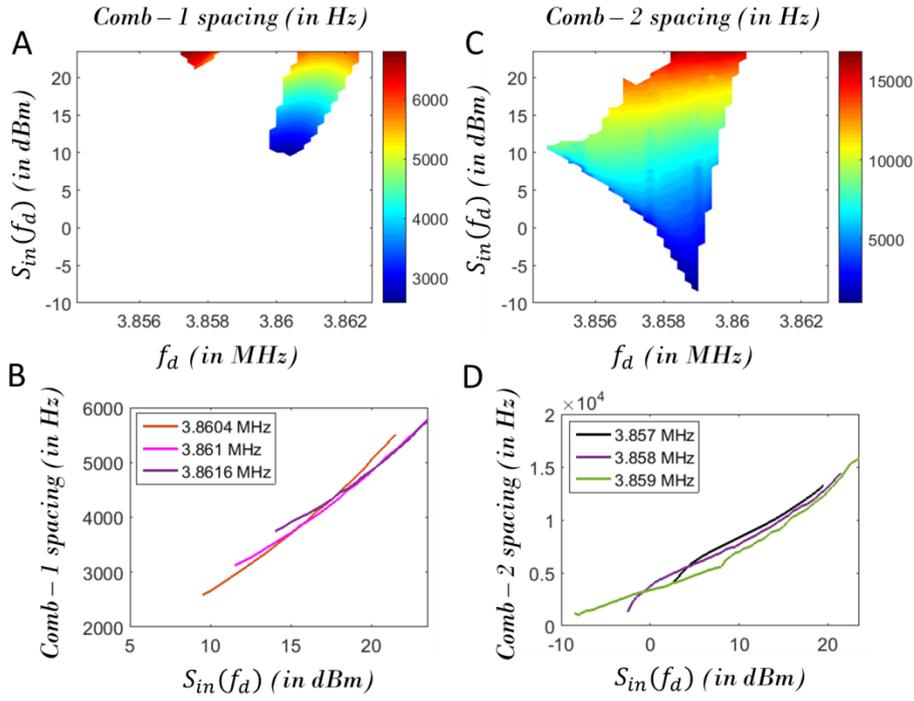

Figure 4: **Spacing of phononic frequency combs.** A & C: The spacing of frequency combs 1 and 2 for a range of drive frequencies $f_d = 3.8544 - 3.8626\ MHz$ and drive power levels $S_{in} = -10 - 23.5\ dBm$. The absence of colour indicates the absence of the respective frequency comb; B: The drive power level dependent spacing of frequency comb 1 for different drive frequencies $f_d = 3.8604, 3.861\ \&\ 3.8616\ MHz$. D: The drive power level dependent spacing of frequency comb 2 for different drive frequencies $f_d = 3.857, 3.858\ \&\ 3.859\ MHz$.

**Supplementary Information**

**Excitation of coupled phononic frequency combs via two-mode parametric three-wave mixing**

Authors: Adarsh Ganesan[1], Cuong Do[1], Ashwin Seshia[1]

1. Nanoscience Centre, University of Cambridge, Cambridge, UK

Supplementary figure S1

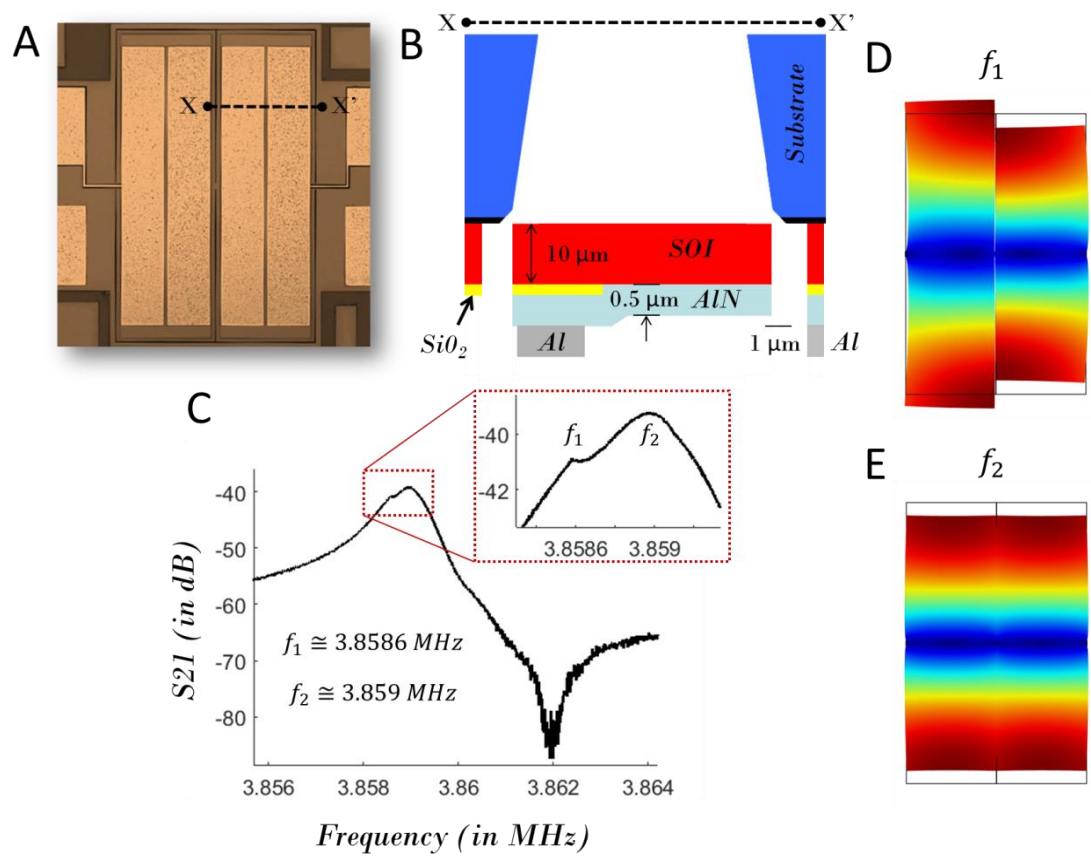

**Figure S1: A**: Coupled Free-free Beams; **B**: 1 μm thick Al electrodes patterned on 0.5 μm thick AlN piezoelectric film which is in-turn patterned on SOI substrate; the 10 μm thick SOI layer is then released through back-side etch to realize mechanical functionality; **C**: The scattering parameter S21 denoting forward transmission gain at the power level of $-10\ dBm$ across the frequency range: 3.856 - 3.864 MHz. This indicates the presence of two neighbouring modes of natural frequencies $f_1$ and $f_2$ respectively; **D** and **E**: The corresponding shapes of these modes.